\documentclass[apj]{emulateapj}
\slugcomment{{\sc Accepted to ApJ Letters:} 2012 September 28}

\usepackage{amsmath}
\usepackage{graphicx}
\usepackage[pdfborder={0 0 1}]{hyperref}
\usepackage{natbib}

\newcommand{\um}{~$\mu$m}

\newcommand{\wcolor}{W1$-$W2}
\newcommand{\afm}{\altaffilmark}

\newcommand{\bessie}{MOO~J2342.0+1301}
\newcommand{\madcows}{MaDCoWS}
\newcommand{\spitzer}{\emph{Spitzer}}

\newcommand\jcap{\ref@jnl{J. Cosmology Astropart. Phys.}}%

\begin{document}

\title{The Massive Distant Clusters of WISE Survey: The First Distant Galaxy Cluster Discovered by WISE}
\author{%
  Daniel P. Gettings\afm{1}, %
  Anthony H. Gonzalez\afm{1}, %
  S. Adam Stanford\afm{2}$^,$\afm{3}, %
  Peter R. M. Eisenhardt\afm{4}, %
  Mark Brodwin\afm{5}, %
  Conor Mancone\afm{1},
  Daniel Stern\afm{4}, %
  Gregory R. Zeimann\afm{3}, %
  Frank J. Masci\afm{6}, %
  Casey Papovich\afm{7}, %
  Ichi Tanaka\afm{8}, %
  Edward L. Wright\afm{9} %
}

\altaffiltext{1}{Department of Astronomy, University of Florida, 211 Bryant Space Center, Gainesville, FL 32611}
\altaffiltext{2}{Institute of Geophysics and Planetary Physics, Lawrence Livermore National Laboratory, Livermore, CA 94550}
\altaffiltext{3}{Department of Physics, University of California, One Shields Avenue, Davis, CA 95616}
\altaffiltext{4}{Jet Propulsion Laboratory, California Institute of Technology, Pasadena, CA 91109}
\altaffiltext{5}{Department of Physics and Astronomy, University of Missouri, 5110 Rockhill Road, Kansas City, MO 64110}
\altaffiltext{6}{Infrared Processing and Analysis Center, Caltech 100-22, Pasadena, CA 91125}
\altaffiltext{7}{George P. and Cynthia Woods Mitchell Institute for Fundamental Physics and Astronomy, and Department of Physics and Astronomy, Texas A\&M University, College Station, TX, 77843-4242, USA}
\altaffiltext{8}{Subaru Telescope, National Astronomical Observatory of Japan, 650 North A'ohoku Place, Hilo, HI 96720}
\altaffiltext{9}{UCLA Astronomy, P.O. Box 951547, Los Angeles, CA 90095-1547}

\begin{abstract}
We present spectroscopic confirmation of a $z=0.99$ galaxy cluster discovered using data from the Wide-field Infrared Survey Explorer (WISE).
This is the first $z\sim1$ cluster candidate from the Massive Distant Clusters of WISE Survey (\madcows) to be confirmed.
It was selected as an overdensity of probable $z\ga1$ sources using a combination of WISE and SDSS-DR8 photometric catalogs.
Deeper follow-up imaging data from Subaru and WIYN reveal the cluster to be a rich system of galaxies, and multi-object spectroscopic observations from Keck confirm five cluster members at $z=0.99$.
The detection and confirmation of this cluster represents a first step towards constructing a uniformly-selected sample of distant, high-mass galaxy clusters over the full extragalactic sky using WISE data.
\end{abstract}
\keywords{galaxies: clusters: individual (\bessie) --- galaxies: distances and redshifts  --- galaxies: evolution}

\section{Introduction}
\label{intro}

Historically, clusters of galaxies have been used as powerful probes of cosmology and galaxy evolution, providing such landmark results as the first evidence for the existence of dark matter \citep{zwicky1937}, demonstration of the importance of environment in galaxy evolution \citep{dressler1980}, and direct proof of the existence of dark matter \citep{clowe2004,clowe2006,bradac2006}. 
The unique leverage provided by galaxy clusters comes primarily from their extreme mass ($M>10^{14}$~M$_{\sun}$) and late-time growth that continues up to the present epoch.

Large-area surveys afford the opportunity to identify well-defined samples of the most massive, rarest galaxy clusters.  
The ROSAT All-Sky Survey, for example, has yielded several notable catalogs of massive X-ray selected galaxy clusters to moderate redshifts \citep[e.g. BCS at $z<0.3$ and MACS at $z<0.7$;][]{ebeling1998,ebeling2001}, while the Sloan Digital Sky Survey (SDSS) has produced large catalogs of nearby clusters covering a wider range of cluster masses \citep[e.g.][ $0.1<z<0.3$]{koester2007}.  
The Planck mission also provides an all-sky catalog of very massive galaxy clusters at $z<1$ \citep{planck2011a,planck2011b}, while the South Pole Telescope (SPT) and Atacama Cosmology Telescope (ACT) provide complementary samples of high-mass clusters reaching to $z>1$ for several thousand square degrees \citep{williamson_2011,marriage2011,reichardt_2012}.
However, there currently exist no surveys capable of identifying massive clusters at $z\ga1$ over the full extragalactic sky.

The most massive clusters in this redshift regime are of particular interest given the recent vigorous debate about whether the few known massive clusters at $z>1$ are consistent with Gaussian primordial density fluctuations \citep{cayon_2010,hoyle_2011,enqvist_2011,williamson_2011,jee2011,hotchkiss_2011}, but a definitive answer remains elusive due to small number statistics.
Moreover, the recent discovery of strong lensing by a galaxy cluster at $z=1.75$ \citep{stanford2012,brodwin2012,gonzalez2012} revives the question of whether the frequency of giant arcs behind clusters at high redshift is consistent with $\Lambda$CDM, a question which can only be comprehensively-addressed with a statistical sample of massive clusters in this regime.

The various wavelength regimes and selection techniques used to construct galaxy cluster samples each offer unique advantages and disadvantages. 
Catalogs selected using the Sunyaev-Zel'dovich (SZ) effect provide high purity samples that are nearly mass-limited due to the detection signal's weak redshift dependence.
These features make SZ surveys ideal for cosmological tests based upon evolution of the cluster mass function.
However, the current generation of SZ catalogs are limited to $z\la1.3$, and at $z>1$ to $M_{200} \ga 3 \times 10^{14}$~M$_{\sun}$ and a few thousand square degrees \citep[e.g.][]{marriage2011,reichardt_2012}.
Similar to the SZ, surveys at X-ray wavelengths provide an estimate of cluster mass immediately from detection.
Current X-ray surveys are able to probe further down the mass function at high-redshift.
Recent searches with XMM-Newton have proven sensitive to clusters out to $z\ga1.6$ for systems with $M_{200}\ga10^{14}$ M$_\sun$ \citep{xmm_clusters_2007,papovich_2010,tanaka_2010,fassbender_2011b}.
The key factor limiting the current X-ray surveys is the area surveyed, roughly an order of magnitude less than the SZ programs.
Complementary to the SZ and X-ray surveys, cluster searches with \spitzer\ based upon detecting galaxy overdensities have, in recent years, provided the greatest reach in redshift and mass sensitivity at high-redshift, extending to $z>2$ and $M\simeq5\times10^{13}$ M$_\sun$ \citep[e.g.][]{iscs_paper,gobat_2011,stanford2012,zeimann_2012}.

The Wide-field Infrared Survey Explorer mission \citep[WISE;][]{wise_paper}, which covers the full sky at 3.4, 4.6, 12, and 22\um, offers the potential to find massive clusters at $z>1$ over the full extragalactic sky.
In this paper we present the first distant galaxy cluster discovered using WISE data, \bessie\ at $z=0.99$. 
The discovery of \bessie\ represents a first step towards constructing an all-sky sample at $1\la z \la 1.4$ as part of the Massive Distant Clusters of WISE Survey (\madcows).
\footnote{ Sources in the \madcows\ catalog are designated as \madcows\ Overdense Objects, which is the origin of the MOO target designation.}


\begin{figure}[t]
 \epsscale{0.90}
 \plotone{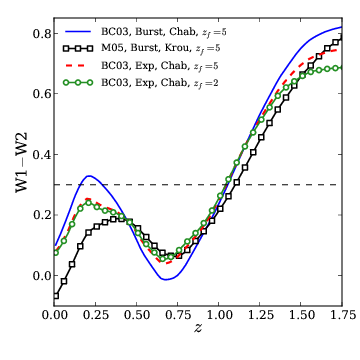}
 \caption{WISE \wcolor\ vs. $z$ for various simulated stellar populations. ``Burst'' models passively evolve after a single starburst lasting 0.1 Gyr. ``Exp'' models have a star formation history of the form $e^{-t/\tau}$ with $\tau=10$ Gyr. Models labeled ``BC03'' come from \citet{bc03} and have a \citet{chabrier2003} initial mass function, while those labeled ``M05'' come from \citet{m05} and have a \citet{kroupa2001} initial mass function. The formation redshift is given by $z_f$. The dashed horizontal line denotes the \wcolor\ cut value used to select \bessie. Models were generated using the EzGal package \citep{ezgal_paper}.}
 \label{fig:model_vs_z}
\end{figure}

\section{WISE Selection of Cluster Candidates}
\label{source_data}

\subsection{WISE Data}
\label{source_data:wise_data}

The key aspect of the WISE mission enabling detection of high redshift galaxy clusters is the excellent photometric sensitivity at 3.4 and 4.6~$\mu$m (W1 and W2, respectively).
The WISE All-Sky Data Release achieves 5$\sigma$ sensitivity limits better than 0.07 and 0.1 mJy in unconfused regions in W1 and W2.\footnote{\url{http://wise2.ipac.caltech.edu/docs/release/allsky/expsup/sec6_3a.html}}
These sensitivities correspond to areas near the ecliptic plane and represent the typical minimum depth for the All-Sky survey.
The effective exposure time of the observations vary significantly with ecliptic latitude due to the survey design, increasing from twelve 7.7s W1 exposures at the ecliptic plane to more than a hundred near the ecliptic poles. 
The WISE All-Sky Source Catalog is based on this imaging data tiled into 18,240 Atlas Images of $\sim2.4$ deg$^2$, from which approximately $5.6\times10^8$ sources were extracted at a 5$\sigma$ detection threshold.
The WISE All-Sky Release data products, including both the Source Catalog and Image Atlas, became public on 2012 March 14 and can be accessed via the NASA/IPAC Infrared Science Archive.\footnote{\url{http://wise2.ipac.caltech.edu/docs/release/allsky}}

\subsection{Sample Definition}
\label{source_data:sample_definition}

Full details of the \madcows\ galaxy cluster search method will be presented in a future paper, but we provide an overview of the essential features here.
The search method that selected \bessie\ was based on the \wcolor\ color, following the approach of \citet{papovich_2008}, who discovered a $z=1.62$ galaxy cluster using \spitzer/IRAC data \citep{papovich_2010}.
This method takes advantage of the fact that galaxy colors in the observed-frame $3-5$~$\mu$m regime become monotonically-redder between $0.75 \la z \la 1.75$, and are largely insensitive to variations in star formation history (see Figure \ref{fig:model_vs_z}).

\begin{figure*}[t]
 \epsscale{0.23}
 \plotone{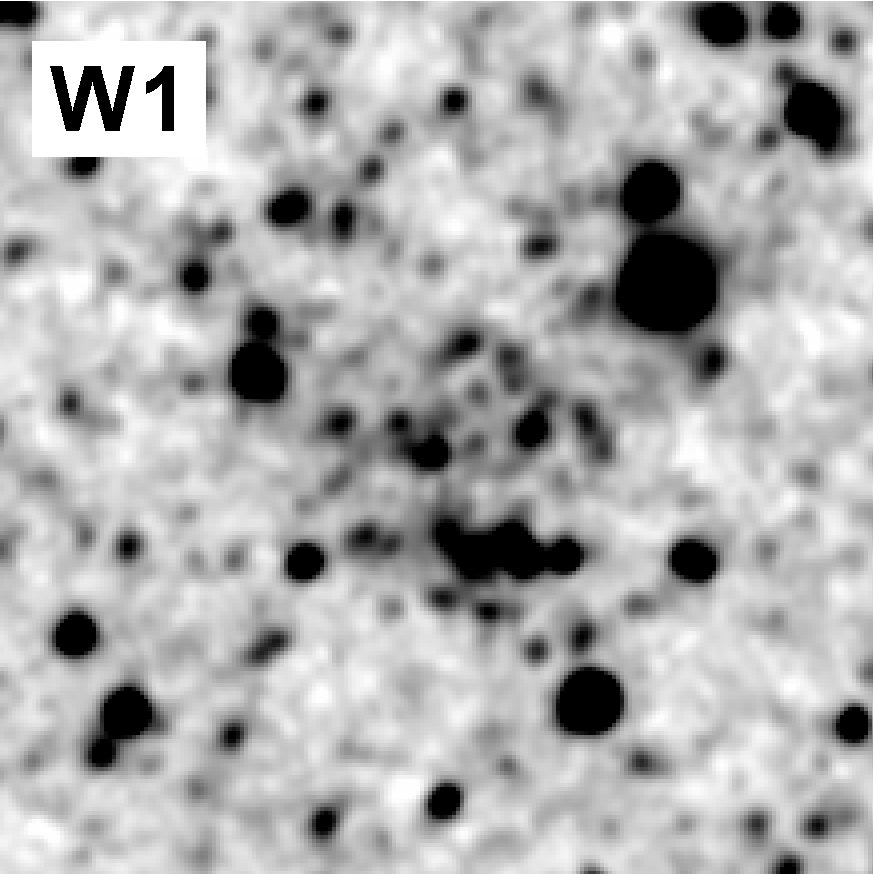}%
 \plotone{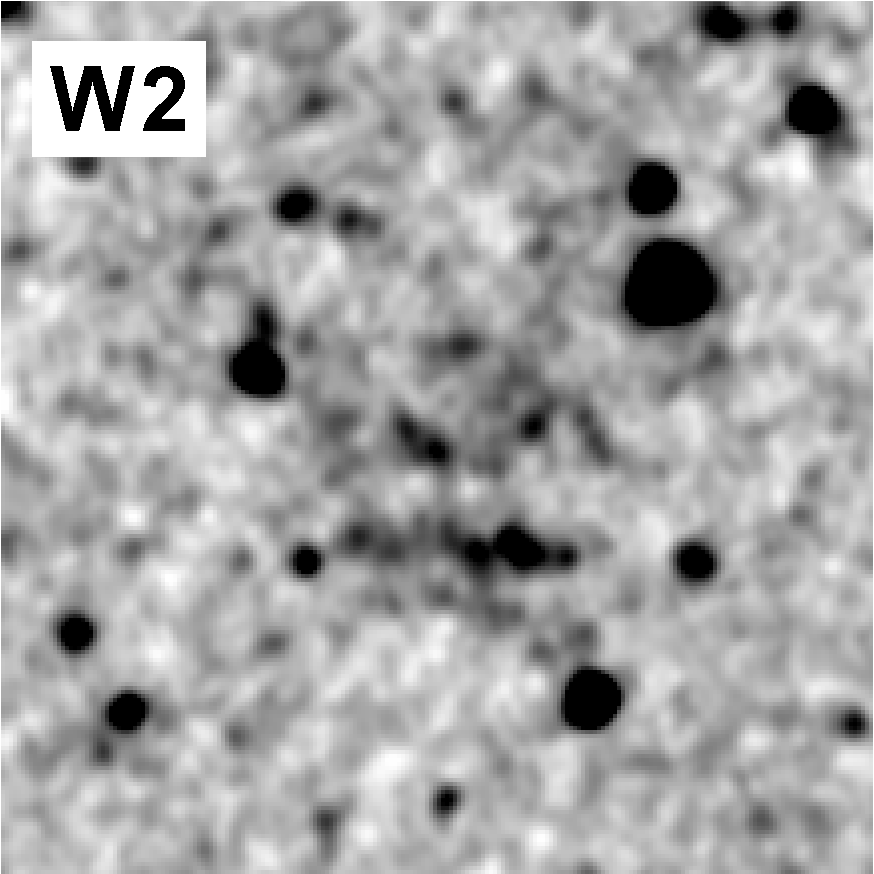}%
 \plotone{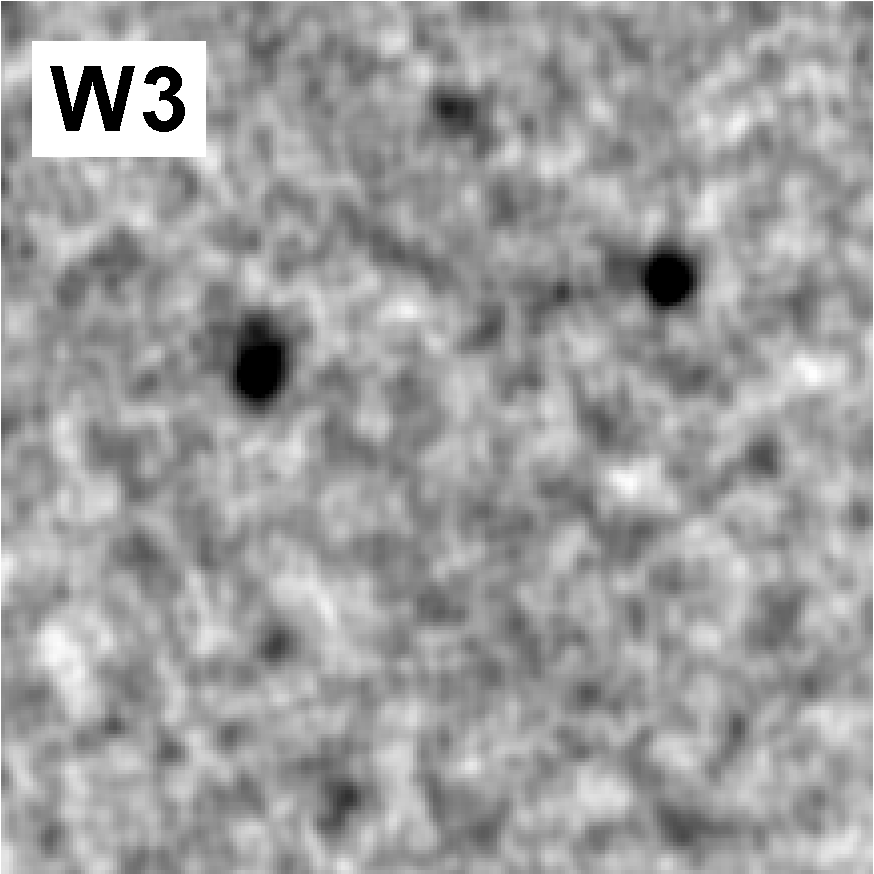}%
 \plotone{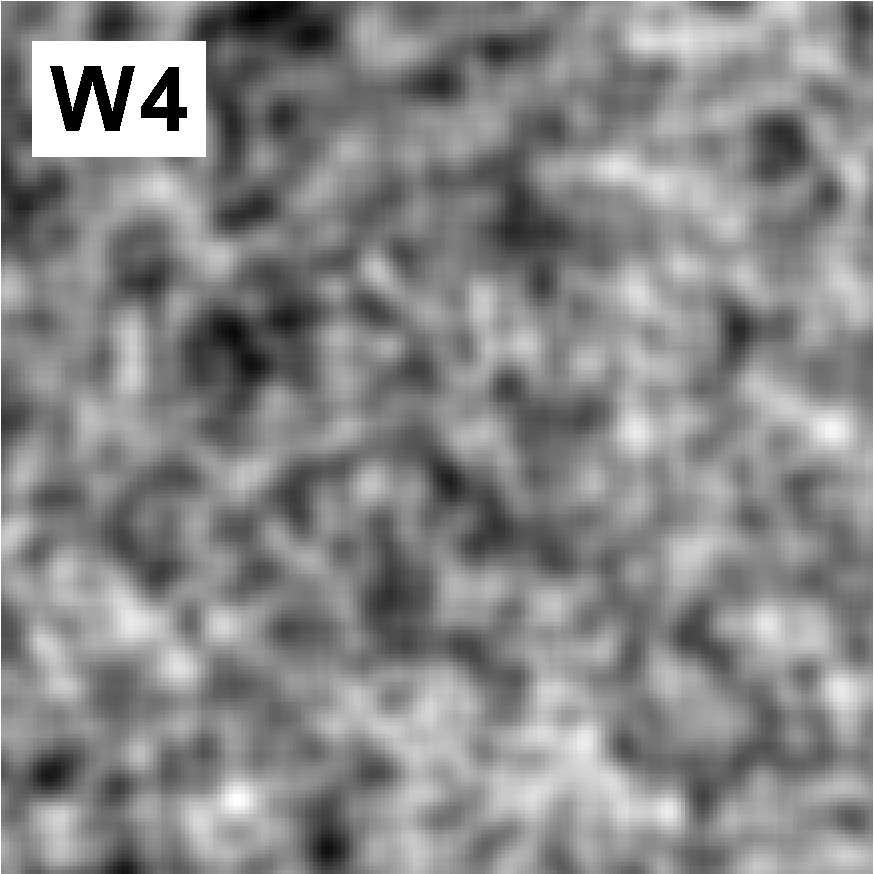}\\
 \vskip 0.2cm
 \epsscale{0.94}
 \plotone{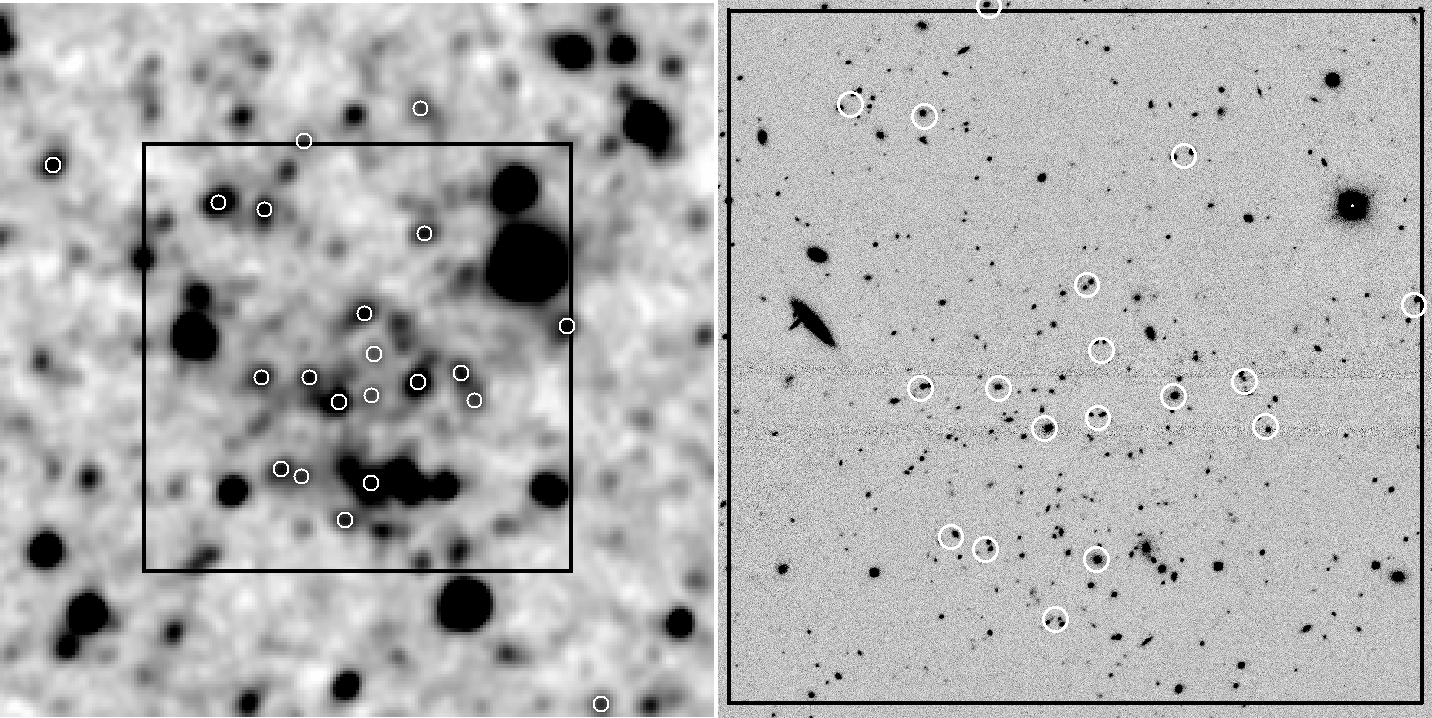}
 \caption{\emph{Top:} $5^\prime\times5^\prime$ images in the four WISE bands, centered at 23h42m05.6s +13d01m29s.  The cluster center is taken from the brightest pixel in its associated wavelet map peak. \emph{Bottom Left:} $5^\prime\times5^\prime$ WISE-W1 image of \bessie. The white circles represent sources identified as probable high-redshift galaxies based upon color and magnitude cuts (see text).  The circles have a diameter of 6.1\arcsec, comparable to the full width half at maximum of the WISE bands W1 and W2 point spread function.  The box denotes a $3^\prime\times3^\prime$ region centered on the detection position. %
   \emph{Bottom Right:} Subaru-MOIRCS $K_s$ image of \bessie.  The box and circles are identical to those in the left panel.  Many of the sources in the WISE imaging are blends which resolve into multiple galaxies in the deeper follow-up imaging. %
 }
 \label{fig:2}
\end{figure*}

After cleaning the WISE catalog of flagged sources and compensating for spatial variations in depth, in the northern hemisphere we matched the WISE extractions to the SDSS-DR8 photometric catalog \citep{dr8_paper}.
For the preliminary search used to identify \bessie, we then rejected sources using a combination of color and optical magnitude cuts, excluding all objects for which W1$-$W2$\;<0.3$, $i-$W1$\;<5$, or $i<21$.
These cuts effectively remove the bulk of the foreground galaxy population at $z<1$.
The remaining sources, consisting predominantly of high-redshift galaxies, were then binned into $\sim 10^{\circ}\times10^{\circ}$ overlapping 2-dimensional density maps with a resolution of $15''$ pix$^{-1}$.
These maps were smoothed with a Gaussian-difference wavelet kernel tuned to enhance structures on scales of $\sim 3'$, corresponding to a physical size of $\sim1.4$~Mpc at $z\sim1$.
From these maps we identified the most statistically significant overdensities for further investigation.
\bessie\ was selected from $\sim 10,000$ deg$^2$ of WISE/SDSS-DR8 overlap and was ranked very highly by both raw overdensity and visual assessments by the \madcows\ team.
Figure \ref{fig:2} shows the cluster in the four WISE bands, centered on the location of the brightest pixel in the associated wavelet map peak.

\section{Follow-Up Observations}
\label{followup}

\subsection{Subaru Imaging}
\label{followup:subaru}
As part of the \madcows\ follow-up program, \bessie\ was imaged using MOIRCS \citep{moircs_paper,moircs_papertwo} on Subaru in $J$ and $K_s$ on UT 2011 September 18 and 19, respectively.
A total of 20 minutes of integration time was obtained in both bands, with $J$ observed in clear conditions and $0\farcs6$ seeing and $K_s$ observed in slightly hazy conditions with $0\farcs8$ seeing.

The MOIRCS data were reduced using a set of IRAF scripts developed by one of the authors \citep{tanaka2011}.   
The reduction procedure follows the standard methodology for the reduction of IR imaging of faint sources.  
After flat fielding, sky subtraction, and cosmic ray removal, the individual frames were corrected to remove geometric distortions, using a solution provided by the Subaru Observatory.  
These corrected frames were then registered and coadded for each detector and filter.  
Astrometric calibration and mosaicing of the chips were performed using the SCAMP \citep{scamp_paper} and SWarp \citep{swarp_paper} software packages, using the Two Micron All Sky Survey as the astrometric reference frame \citep[2MASS;][]{skrutskie2006}.
The data were also photometrically calibrated using 2MASS. 
The $K_s$ image is shown in Figure \ref{fig:2}.

\subsection{WIYN Imaging}
\label{followup:wiyn}

\begin{figure*}[t]
  \epsscale{0.45}
  \plotone{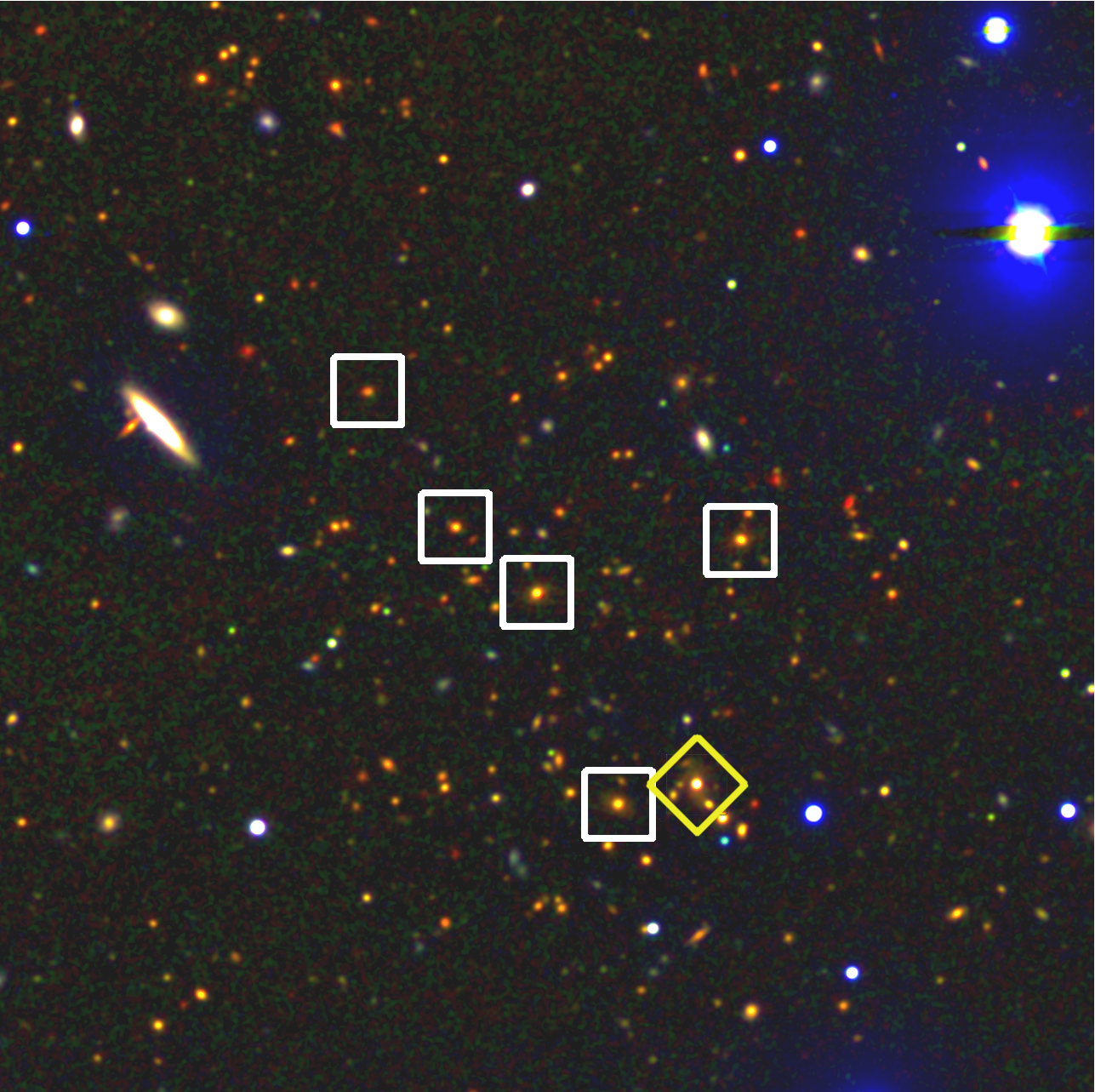}
  \epsscale{0.47}
  \plotone{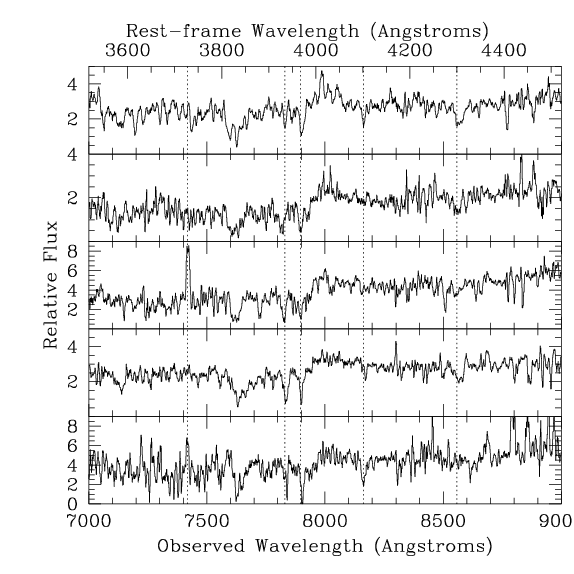}
 \caption{ \emph{Left:} A WIYN-$i$ and Subaru $J K_s$ 3-color image of the central $3^\prime\times3^\prime$ region of \bessie.  Boxes indicate spectroscopically confirmed cluster members.  The diamond denotes an object with a redshift that is not at the cluster redshift. Of the six galaxies targeted in the LRIS mask which are within 500~kpc of the nominal cluster center, five were confirmed to be cluster members. %
 \emph{Right:} LRIS spectra of the five member galaxies.  The rest-frame wavelength at $z=0.99$ is shown along the top. The vertical dotted lines mark the positions of the following features (from left to right): [\ion{O}{2}]$\lambda$3727, Ca H+K, H$\delta$, and the G-band.
 }
 \label{fig:specz}
\end{figure*}

We obtained Sloan $i$-band imaging from the Mini-Mosaic instrument \citep{minimo_paper} on WIYN during the night of UT 2011 November 18.  
A total of 60 minutes of exposure was obtained in $0\farcs9$ seeing using three dithered 20 minute integrations.
The data were processed using standard reduction procedures in IRAF and astrometrically calibrated to the USNO-B1.0 catalog \citep{usnob_paper} using SCAMP and SWarp.
Photometric calibration of final, stacked images was performed using SDSS-DR8 stars in the field.
A color composite of the $i$, $J$ and $K_s$ images is shown in Figure \ref{fig:specz}.

\subsection{Keck Spectroscopy}
\label{followup:keck}

Based on the Subaru imaging results, a slit-mask was designed to be used with the Low-Resolution Imaging Spectrometer \citep[LRIS;][]{lris_paper} at the Keck Observatory to obtain redshifts of potential cluster galaxies.  
Sources for the mask were selected using $J - K_s$ colors, $K_s$ magnitudes, and distance from the nominal cluster center, with final selection also accounting for slit-positioning constraints.  
Spectra were obtained on UT 2011 October 22 using $1\farcs1 \times 10''$ slitlets, the G400/8500 grating on the red side, the D680 dichroic, and G300/5000 grism on the blue side.   
Four 1200~s exposures were obtained in mostly clear conditions with $0\farcs6$ seeing.  

The LRIS spectra were split into slitlets which were separately reduced using standard long-slit procedures in IRAF.    
The relative spectral response was calibrated via longslit observations of Wolf 1346 and Hiltner 600.  
Redshifts were determined by visual inspection using prominent spectral features including the 4000 Angstrom break, the Ca H+K absorption lines, and the [\ion{O}{2}] $\lambda 3727$ emission line.  

\section{Results}
\label{results}

\subsection{Redshifts}
\label{results:redshifts}

The Keck spectroscopy yielded high-quality redshifts for 12 galaxies.  
Geometrical constraints on the slit placement allowed observations of only six galaxies within a projected distance of 500 kpc from the nominal cluster center.
Five were confirmed to be cluster members, with redshifts lying within $\Delta z = 0.010$, or $\pm 750$ km s$^{-1}$, of $z=0.987$,  the mean cluster redshift.
In Table \ref{spec_table} we provide coordinates, redshifts, identifying features, and a quality assessment for each redshift.
All but one of the confirmed cluster members were identified based only on absorption lines and continuum breaks, rather than emission lines, a feature characteristic of the members of mature galaxy clusters. 
The spectra of the five confirmed cluster members are shown in Figure \ref{fig:specz}.

\subsection{Color--Magnitude Diagram}
\label{results:cmd}

\begin{figure*}[t]
  \epsscale{0.90}
  \plotone{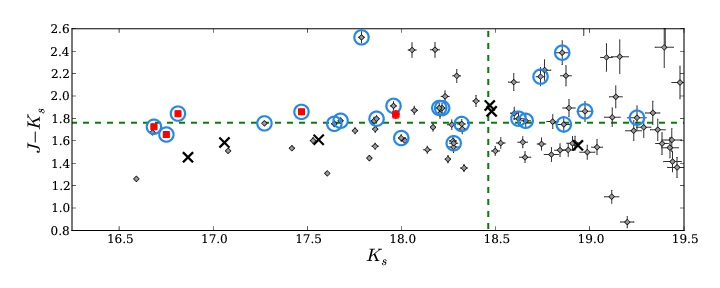}
  \caption{$(J-K_s)$ vs. $K_s$ color-magnitude diagram for \bessie. Gray diamonds denote galaxies that lie within a 500~kpc projected radius of the cluster center with $i-J>1.38$.  Black squares (red in the online version) are spectroscopically confirmed cluster members.  Black ``X'' points represent non-member galaxies with spectroscopic redshifts. Large circles (blue in the online version) indicate all objects found within 3.05\arcsec\ of the WISE sources that contributed to the detection signal (see Figure \ref{fig:2}). The vertical and horizontal dashed lines show the expected $K_s$ magnitude and $J-K_s$ color for an $L^*$ galaxy formed at $z_f=5$ and observed at $z=0.99$. All magnitudes are on the Vega system.}
 \label{fig:cmd}
\end{figure*}

We use imaging data from Subaru and WIYN to construct the $J-K_s$ color-magnitude diagram (CMD) for \bessie.
Photometry was obtained using SExtractor version 2.5.0 \citep{sextractor_paper} in dual-image mode with $K_s$ used for source detection.
We present the resulting CMD in Figure \ref{fig:cmd}.
Light grey diamonds show all galaxies within a projected 500~kpc radius of the cluster center that have $i-J>1.38$, the color given by a \citet{bc03} exponential model with $\tau=10$~Gyr, formed at $z_f=5$ and observed at $z=0.99$.
Spectroscopically confirmed members and non-members are marked with black boxes (red in the online version) and black crosses, respectively.
Additionally, we denote with large circles (blue in the online version) all $K_s$-selected objects that lie within the footprint of the $\sim$6.1\arcsec~ WISE PSF for each WISE source that satisfies the color and magnitude cuts described in Section \ref{source_data:sample_definition} (see also Figure \ref{fig:2}). It is these WISE sources, the majority of which are composed of multiple blended components, that contributed to the detection signal for the cluster.
We include lines showing the expected $J-K_s$ color (horizontal) and apparent $K_s$ magnitude (vertical) of an $L^*$ galaxy at $z=0.99$, using a \citet{bc03} simple stellar population model with solar metallicity and a \citet{chabrier2003} initial mass function, formed at $z_f=5$.
The red sequence in this cluster is well-populated, with both the spectroscopic members and galaxies corresponding to the WISE sources contributing to detection of the cluster lying near the expected $z=0.99$ color.

\section{Discussion}
\label{discussion}

The discovery of \bessie\ represents the first result from \madcows, whose goal is to conduct a wide-area search for such massive clusters at $1\la z\la 1.4$.
This program is designed to be complementary to previous, shallower all-sky cluster surveys, enabling future investigations of the most extreme clusters at this epoch. 
Forthcoming papers will present details for the full \madcows\ program, including the ongoing followup program and characterization of the cluster selection function.

Finally, we note that this cluster was found using the WISE All-Sky Data Release, which includes only data obtained during the cryogenic mission. 
Data continued to be collected at 3.4~\um\ and 4.6~\um\ after exhaustion of the WISE cryogen supply as part of the NEOWISE \citep{neowise_paper} survey of the asteroid belt.
However, the asteroid search required only that individual exposures be processed, so the multiple observations of inertially fixed sources from this phase of the mission have not been combined.
Fully processing and combining this post-cryogenic data with the cryogenic data in these bands, which has now been funded by NASA, would double the all-sky coverage from WISE in W1 and W2, improving the fidelity and redshift reach of WISE for discovering the most massive, distant galaxy clusters.

\begin{deluxetable*}{llccccl}[!]
\tablewidth{0pt}
\tablecaption{Spectroscopic Results}
\tablehead{ \colhead{$\alpha$ (J2000)}& \colhead{$\delta$ (J2000)} & \colhead{$J$} & \colhead{$K_s$} & \colhead{$z$} & \colhead{ Quality} & \colhead{Features}   }
\startdata
\multicolumn{7}{c}{Members}\\
23:42:04.80 & +13:00:50.0 & 18.40 & 16.75 &   0.989 &  A &  [\ion{O}{2}], D4000, Ca H+K \\
23:42:03.43 & +13:01:32.7 & 18.66 & 16.81 &   0.993 &  B &  D4000, Ca H+K \\
23:42:05.67 & +13:01:24.2 & 18.37 & 16.69 &   0.983 &  A &  D4000, Ca H+K \\
23:42:06.57 & +13:01:34.9 & 19.33 & 17.47 &   0.987 &  B &  D4000, Ca H+K \\
23:42:07.56 & +13:01:56.8 & 19.81 & 17.97 &   0.985 &  A &  D4000, H$\delta$ \\
\multicolumn{7}{c}{Non-Members}\\
23:42:03.93 & +13:00:53.2 & 17.65 & 16.00 &  0.784 & B & Ca H+K \\	
23:41:57.98 & +13:00:51.0 & 19.17 & 17.56 &  1.043 & B & [\ion{O}{2}] \\	
23:41:58.15 & +12:59:17.1 & 18.66 & 17.06 &  0.777 & A & Ca H+K, D4000	 \\	
23:42:08.86 & +13:02:47.9 & 20.38 & 18.47 &  1.210  & B & D4000 \\	
23:41:59.87 & +12:59:14.9 & 20.47 & 18.94 &  1.172 & A & [\ion{O}{2}] \\	
23:41:59.28 & +12:58:09.6 & 20.34 & 18.48 &  0.622 & A & [\ion{O}{2}], H$\beta$  \\		
23:42:00.65 & +12:58:16.2 & 18.33 & 16.87 &  0.249 & A & H$\alpha$, [\ion{S}{2}] \\	
\enddata
\tablecomments{Coordinates, redshifts, identifying features, and quality assessments of each redshift obtained from LRIS observations on Keck I. The typical redshift uncertainty is $\delta z \simeq 0.001$. Quality flag ``A'' signifies a robust redshift determination, typically relying upon multiple emission or absorption features. Quality flag ``B'' signifies a redshift determination that is less certain, but still unambiguous with at least two features. Magnitudes are on the Vega system. %
}
\label{spec_table}
\end{deluxetable*}

\acknowledgements
The authors thank the anonymous referee whose comments improved the quality of the manuscript. 
This publication makes use of data products from the Wide-field Infrared Survey Explorer, which is a joint project of the University of California, Los Angeles and the Jet Propulsion Laboratory/California Institute of Technology, funded by the National Aeronautics and Space Administration (NASA).
D. P. G. and A. H. G. acknowledge support for this research from the NASA Astrophysics Data Analysis Program (ADAP) through grant NNX12AE15G. 
Some of the data presented herein were obtained at the W.M. Keck Observatory, which is operated as a scientific partnership among the California Institute of Technology, the University of California and the National Aeronautics and Space Administration. The Observatory was made possible by the generous financial support of the W.M. Keck Foundation. 
Based in part on data collected at Subaru Telescope, which is operated by the National Astronomical Observatory of Japan.
D. P. G. was a Visiting Astronomer, Kitt Peak National Observatory, National Optical Astronomy Observatory, which is operated by the Association of Universities for Research in Astronomy (AURA) under cooperative agreement with the National Science Foundation.
The WIYN Observatory is a joint facility of the University of Wisconsin-Madison, Indiana University, Yale University, and the National Optical Astronomy Observatory.
This research has made use of the NASA/ IPAC Infrared Science Archive, which is operated by the Jet Propulsion Laboratory, California Institute of Technology, under contract with the National Aeronautics and Space Administration.

\newpage
\newpage

\bibliographystyle{apj}

\end{document}